\documentclass[a4paper,11pt]{article}
\pdfoutput=1 

\usepackage{jheppub} 

\usepackage[T1]{fontenc} 
\usepackage{bm,xurl}
\usepackage{bm,slashed}
\usepackage{ragged2e}

\title{\boldmath On the possibility of differential-algebraic elimination of the spinor field from the Maxwell--Dirac electrodynamics}

\author[a]{A. Akhmeteli}

\affiliation[a]{LTASolid Inc.,\\9797 Meadowglen Ln 506, Houston 77042, USA}

\emailAdd{akhmeteli@ltasolid.com}

\abstract{We investigate whether the spinor field can be differential-algebraically eliminated from the Maxwell–Dirac equations in a particular gauge. To this end, we construct a generic truncated power-series solution and linearize the prolonged system of the Maxwell--Dirac equations about this solution. We then analyze the ranks of the coefficient matrices associated with the linearized system. Our results indicate that, generically, the spinor components are uniquely determined by the electromagnetic field and its derivatives. Furthermore,  the fourth-order time derivatives of the components of the electromagnetic four-potential are uniquely determined by derivatives of the lower order with respect to time. These findings strongly suggest that the spinor field can be differential-algebraically eliminated, and the resulting equations describe independent evolution of the electromagnetic field, i.e., a Cauchy problem can be formulated in terms of the electromagnetic variables alone.}

\begin{document} 
\maketitle
\flushbottom

\section{Introduction}
\label{sec:intro}
In 1951, Dirac ~\cite{Diracmain} showed that the standard Lagrangian of the free electromagnetic field with a certain gauge-fixing constraint yields self-consistent equations of motion describing both the electromagnetic field in a particular gauge and classical particles moving under the Lorentz force. In a comment on Dirac's paper, Schrödinger~\cite{Schroed} noted that the Maxwell--Klein--Gordon electrodynamics in the unitary gauge ~\cite{Itzykson}, where the scalar matter field is real, has the same modified Maxwell equations for the electromagnetic field, up to a gauge, as in ~\cite{Diracmain}. It was later shown that in this case the matter field can be algebraically eliminated from the equations, and the resulting modified Maxwell equations describe independent evolution of the electromagnetic field ~\cite{Akhmeteli-EPJC}. Let us also note that effects of restriction of gauge symmetries were discussed in  ~\cite{KamBar,KamVen}.

It is therefore natural to ask whether a similar result can be obtained for the more realistic Maxwell--Dirac electrodynamics (spinor electrodynamics). A positive answer to this question was given in ~\cite{Akhmeteli-EPJC}, but only at the cost of introducing a complex four-potential of the electromagnetic field. The present work takes a major step towards removing this shortcoming.

More specifically, we ask whether the spinor field can be algebraically or differential-algebraically eliminated from the equations of the Maxwell--Dirac electrodynamics. It was shown in ~\cite{Akhmeteli-JMP} that, generically, three of the four spinor components can be algebraically eliminated from the Dirac equation in the electromagnetic field, while the remaining component can be made real by a gauge transformation, at least locally. However, eliminating this last component from the full Maxwell--Dirac system is much more difficult. In principle, methods of differential algebra could be used for this purpose, but the required computational resources appear to be impractically large. In particular, the author attempted to use the Maple \textit{DifferentialThomas} package  ~\cite{Gerdt,difthom} and E. Mansfield's Indiff package ~\cite{Mansfield,Indiff}, but without success. The present paper therefore adopts a different approach.

\section{Evidence for the possibility of eliminating the spinor field from the Maxwell--Dirac electrodynamics}
\label{sec:1}
The Dirac equation in electromagnetic field $A_\mu$ is
\begin{equation}\label{eq:pr25qr22}
(i\slashed{\partial}-\slashed{A})\psi=\psi,
\end{equation}
where the Feynman slash notation is used ($\slashed{A}=A_\mu\gamma^\mu$). Units are chosen so that $\hbar=c=m=1$, and the electric charge $e$ is absorbed into $A_\mu$. We use the chiral representation of the $\gamma$-matrices ~\cite{Itzykson}:
\begin{equation}\label{eq:d1qr22}
\gamma^0=\left( \begin{array}{cc}
0 & -I\\
-I & 0 \end{array} \right),\gamma^i=\left( \begin{array}{cc}
0 & \sigma^i \\
-\sigma^i & 0 \end{array} \right),
\end{equation}
where the index $i$ runs from 1 to 3, $I$ is the 2-dimensional identity matrix, and $\sigma^i$ are the Pauli matrices.

In terms of the four components of the Dirac spinor, the Dirac equation takes the form
\begin{eqnarray}\label{eq:d3qr22}
(A^0+A^3)\psi_3+(A^1-i A^2)\psi_4+
i(\psi_{3,3}-i\psi_{4,2}+\psi_{4,1}-\psi_{3,0})=\psi_1,
\end{eqnarray}
\begin{eqnarray}\label{eq:d4qr22}
(A^1+i A^2)\psi_3+(A^0-A^3)\psi_4-
i(\psi_{4,3}-i\psi_{3,2}-\psi_{3,1}+\psi_{4,0})=\psi_2,
\end{eqnarray}
\begin{eqnarray}\label{eq:d5qr22}
(A^0-A^3)\psi_1-(A^1-i A^2)\psi_2-
i(\psi_{1,3}-i\psi_{2,2}+\psi_{2,1}+\psi_{1,0})=\psi_3,
\end{eqnarray}
\begin{eqnarray}\label{eq:d6qr22}
\nonumber
-(A^1+i A^2)\psi_1+(A^0+A^3)\psi_2+
\\
i\psi_{2,3}+\psi_{1,2}-i(\psi_{1,1}+\psi_{2,0})=\psi_4.
\end{eqnarray}
We impose the gauge condition that $\psi_1$ is real.

The Maxwell equations are
\begin{equation}\label{eq:pr26qr24}
\Box A_\mu-A^\nu_{,\nu\mu}=e^2\bar{\psi}\gamma_\mu\psi,
\end{equation}
where the d'Alembertian is
\begin{equation}\label{eq:pr26qr24a}
\Box=\partial^\mu\partial_\mu.
\end{equation}

The strategy used in this work is as follows. First, we pose the Cauchy problem and compute prolongations of the equations by derivation. Ideally, one would like to express the spinor components and their derivatives directly in terms of the components of the electromagnetic field and their derivatives. However, solving the resulting large system of polynomial equations does not appear to be feasible in practice.

We therefore proceed indirectly. A particular truncated power-series solution is constructed. It is specified by the values of derivatives of the real dependent variables, i.e., the components of the spinor and electromagnetic fields, at a fixed point. Some of these derivatives are chosen randomly as rational-valued initial data, while the remaining derivatives are determined from the Maxwell--Dirac equations and their prolongations. The truncated series is an approximation to the solution, but it is a part of an infinite series that is a precise solution, so the truncated series provides complete information on the derivatives of up to a certain order.

Next, we linearize the Maxwell--Dirac equations and their prolongations about this particular solution. We then consider the resulting linear system for the variations of the spinor components and their derivatives, while the components of the electromagnetic field and their derivatives are fixed at the values provided by the particular solution. We thus investigate solutions that are infinitesimally close to the chosen power series.

The linearized system always has the trivial solution. If we can show that the variation of the spinor component $\psi_\alpha$ is uniquely determined, then the corresponding value $\psi_\alpha$ in the particular solution is isolated. This would strongly suggest that, generically, the values of the electromagnetic field and their derivatives uniquely determine $\psi_\alpha$ at any point.

After fixing the gauge, one obtains a system of partial differential equations (PDE) for eleven real dependent variables:
\begin{eqnarray}\label{eq:pr26qr24b}
\nonumber
A^0, A^1, A^2, A^3,\psi_{1r}=\mathrm{Re}(\psi_1),\psi_{2r}=\mathrm{Re}(\psi_2),
\\
\psi_{2i}=\mathrm{Im}(\psi_2),
\psi_{3r}=\mathrm{Re}(\psi_3), \psi_{3i}=\mathrm{Im}(\psi_3),
\psi_{4r}=\mathrm{Re}(\psi_4), \psi_{4i}=\mathrm{Im}(\psi_4).
\end{eqnarray}
To analyze this system, we used the Maple \textit{DifferentialThomas} package  ~\cite{Gerdt} to compute an equivalent so-called simple and integrable system of equations; the corresponding Maple code is provided at ~\cite{DT2}. The dependent variables were ordered as
\begin{equation}\label{eq:pr26qr24c}
[A^0,\psi_{4i},\psi_{4r},\psi_{3i},\psi_{3r},\psi_{2i},\psi_{2r},A^3,A^2,A^1,\psi_{1r}],
\end{equation}
the independent variables were taken to be
\begin{equation}\label{eq:pr26qr24d}
[x^0,x^1,x^2,x^3],
\end{equation}
and the ranking was chosen to be \textit{DegRevLex}. The ranking \textit{EliminateFunction} was also tested, but it required prohibitively large computational resources. The resulting simple and integrable system consists of thirteen PDEs whose leaders, i.e., the highest-ranking variables appearing in the corresponding polynomials, are
\begin{eqnarray}\label{eq:pr26qr24e}
\nonumber
A^0_{,1000},\psi_{4i,1000},\psi_{4r,1000},\psi_{3i,1000},\psi_{3r,1000}, 
\\
\psi_{2i,1000},\psi_{2i,0100},\psi_{2r,1000},A^3_{,2000},
A^2_{,2000},A^1_{,2000},A^1_{,1100},\psi_{1r,2000},
\end{eqnarray}
 where, for example,
 \begin{equation}\label{eq:pr26qr24f}
A^1_{,1100}=\frac{\partial^2 A^1}{\partial x^0\partial x^1}.
\end{equation}

For the equations with leaders $A^1_{,1100}$, $\psi_{2i,0100}$, the independent variables admissible for prolongation, i.e., those with respect to which one can repeatedly differentiate the equation, are $x^1,x^2,x^3$. For all other equations, prolongations with respect to each of $x^0,x^1,x^2,x^3$ are admissible.

This information was then used to specify the rational initial data for the Cauchy problem. Rational values are needed as later we compute matrix ranks exactly. With our choice of units,  the elementary charge squared satisfies $e^2=4\pi\alpha$, where $\alpha$ is the fine-structure constant. Accordingly, we replaced $e^2$ in the equations by the rational approximation $221/2410$. The derivatives appearing in the initial data were chosen as rational numbers with random signs and with random integer numerators and denominators between 1 and 9 inclusive.

We compute prolongations of the original Maxwell--Dirac equations rather than of the equivalent simple and integrable system. The reason is that, in the original system, the polynomials in the dependent variables and their derivatives have degree at most two, which makes the prolongations easier to compute.

The initial data were specified for the following variables and derivatives:
\begin{eqnarray}\label{eq:pr26qr24f2}
\nonumber
A^0_{,0 i_1 i_2 i_3},A^1_{,0 i_1 i_2 i_3},\psi_{4i,0 i_1 i_2 i_3},\psi_{4r,0 i_1 i_2 i_3},\psi_{3i,0 i_1 i_2 i_3},\psi_{3r,0 i_1 i_2 i_3},\psi_{2r,0 i_1 i_2 i_3},
\\
\psi_{1r,0 i_1 i_2 i_3},A^3_{,0 i_1 i_2 i_3},
A^2_{,0 i_1 i_2 i_3},A^3_{,1 i_1 i_2 i_3},A^2_{,1 i_1 i_2 i_3},\psi_{2i,0 0 i_2 i_3},A^1_{,10 i_2 i_3},
\end{eqnarray}
where $i_1,i_2,i_3$ are nonnegative integers. Higher-order derivatives, for example, derivatives of higher order with respect to time $x^0$, were then computed from the original equations and their prolongations. Only the original equations and prolongations containing derivatives of total order not greater than five were used. In this way, a particular truncated power-series solution was built.

We now ask whether there exist solutions infinitesimally close to this particular solution that have the same values of the electromagnetic field $A^{\mu}$ and its derivatives, but a different value of, say, $\psi_1$. To answer this question, we linearize the system about the particular solution. If the values of the electromagnetic and spinor fields in the particular solution are
\begin{equation}\label{eq:pr26qr24g}
A^{0(0)}_{,i_0 i_1 i_2 i_3},\psi^{(0)}_{1r,i_0 i_1 i_2 i_3},
\end{equation}
 and so on, we consider a perturbed solution of the form
\begin{equation}\label{eq:pr26qr24h}
 A^{0(0)}_{,i_0 i_1 i_2 i_3},\psi^{(0)}_{1r,i_0 i_1 i_2 i_3}+\delta\cdot\psi^{(1)}_{1r,i_0 i_1 i_2 i_3},
\end{equation}
and so on. Thus, the electromagnetic field and its derivatives are fixed at the values of the particular solution, whereas the spinor components and their derivatives acquire infinitesimal variations. Substituting this ansatz into the original equations and their prolongations and retaining only  terms of the first order in $\delta$, we obtain a system of linear equations for
\begin{equation}\label{eq:pr26qr24i}
 \psi^{(1)}_{1r,i_0 i_1 i_2 i_3}
\end{equation}
and the analogous variations of the other spinor variables.

The linearized system  always has the trivial solution, because it is obtained by linearization about a particular solution. The question is whether it also admits nontrivial solutions with, for example,
\begin{equation}\label{eq:pr26qr24j}
 \psi^{(1)}_{1r,0 0 0 0}\neq 0.
\end{equation}
To answer this, we consider the coefficient matrix of the linearized system and compare its rank with the rank of the matrix obtained by deleting the column corresponding to $\psi^{(1)}_{1r,0 0 0 0}$. As shown in the Appendix, if these two ranks differ by $1$, then $\psi^{(1)}_{1r,0 0 0 0}$ is uniquely determined and therefore must be zero. In the present case, the computed ranks are $669$ and $668$, respectively. We therefore conclude that the components of the electromagnetic field and their derivatives determine an isolated value of $\psi_{1r}$.

Since the coefficients of the linearized system are polynomial functions of the jet variables of the background solution, the rank condition is equivalent to the nonvanishing of certain polynomial minors. Therefore, if the condition holds for one particular rational jet, it also holds for algebraically generic jets, i.e., outside a special set defined by polynomial relations among the jet variables. Thus, the observed rank behavior reflects the generic structure of the prolonged Maxwell–Dirac system rather than a special feature of the chosen numerical example.

The same result was obtained for all other spinor components, but not for their derivatives. Again, only the original equations and prolongations containing derivatives of total order not greater than five were used. If the maximal total order is reduced to four, the relevant matrix ranks coincide, and no analogous conclusion can be drawn. Additional prolongations are needed in order to obtain similar results for derivatives of the spinor components.

It is also essential to determine whether the equations resulting from elimination of the spinor variables describe independent evolution of the electromagnetic field, i.e., whether a Cauchy problem can be formulated in terms of the electromagnetic variables only. To study this, we again linearized the Maxwell--Dirac equations and their prolongations including derivatives of total order not greater than five, but this time we did not fix all derivatives of $A^{\mu}$. Instead, we fixed only those derivatives whose order with respect to time was less than four, while their total order could still be higher because of differentiation with respect to spatial coordinates. We then considered the coefficient matrix of the resulting linearized system and compared its rank with the rank of the matrix obtained by deleting the columns corresponding to the fourth-order time derivatives
\begin{equation}\label{eq:pr26qr24k}
 A^{\mu(1)}_{,4000}.
\end{equation}
As shown in Appendix, if these two ranks differ by $4$, then the variations of the derivatives (\ref{eq:pr26qr24k}) are uniquely determined and therefore must vanish. The computed ranks are $685$ and $681$. This indicates that a Cauchy problem can be formulated in terms of the electromagnetic variables only. There may still be some constraints on the initial data, but the resulting system is consistent, since it is derived from the consistent Maxwell--Dirac system.

One may question the use of a rational approximation to the dimensionless charge, since in principle the rank computations could yield different results for the exact value of the fine-structure constant. However, such a possibility would imply that the present approach could in effect yield a theoretical determination of the fine-structure constant, a striking consequence that seems highly unlikely. 

The computation does not constitute a proof of global differential-algebraic elimination. Rather, it shows that, for a generic truncated jet of sufficiently high order, the linearized systems for the spinor variations and for the variations of the fourth-order time derivatives of $A^{\mu}$ have trivial kernels. This strongly suggests that the spinor is locally determined by the electromagnetic field and its derivatives, and the equations resulting from elimination of the spinor variables describe independent evolution of the electromagnetic field.
\section{Conclusion}
\label{sec:2}
Elimination of the spinor field from the Maxwell--Dirac equations by currently available differential-algebraic does not appear to be computationally feasible, since it would require handling of a large polynomial system. In this work, the Maxwell--Dirac equations in a fixed gauge, together with their prolongations, are linearized about a random truncated power-series solution. The computations show that, when the electromagnetic field is fixed, the resulting linear system for the variations of the spinor components has only the trivial solution. This strongly suggests that, generically, the set of spinor solutions of the Maxwell--Dirac equations for a fixed electromagnetic field is null-dimensional. Accordingly, it is highly plausible that the spinor field can be expressed in terms of the components of the electromagnetic field and their derivatives. Computations based on a similar approach further indicate that the resulting equations for the electromagnetic field describe its independent evolution. 

These results are important in their own right and may also be relevant to the interpretation of quantum theory. For example, they suggest that in the de Broglie--Bohm interpretation (see, e.g., ~\cite{Holland}) the electromagnetic field can play the role of the guiding field (cf. ~\cite{Akhmeteli-IJQI}) . The results are also relevant to the derivation of the equations of motion in the plasma-like models of elementary particles described in ~\cite{Akhmeteliqr}.

\appendix
\section{Appendix}
\label{sec:3}
Let us consider a linear system of equations
\begin{equation}\label{eq:att1k}
Ax=0,
\end{equation}
where $A$ is an $m\times n$ matrix. Let $J\subset \{1,\ldots,n\}$ be a set of column indices, with $|J|=k$. Denote by $A_{-J}$ the matrix obtained from $A$ by deleting all columns with indices from $J$, and write
\begin{equation}\label{eq:att1k2}
x=(x_J,x_{-J}),
\end{equation}
where $x_J$ is the vector of components of $x$ with indices in $J$.

Let $A_J$ be the $m\times k$ submatrix of $A$ formed by the columns with indices in $J$. Then
\begin{equation}\label{eq:att1k3}
A=[A_{-J}\mid A_J].
\end{equation}

Let us prove that if
\begin{equation}\label{eq:rankcondk}
\operatorname{rank}(A)=\operatorname{rank}(A_{-J})+k,
\end{equation}
then equation (\ref{eq:att1k}) uniquely determines $x_J$, i.e.,
\begin{equation}\label{eq:kernelcondk}
\forall v\in\ker(A),\quad v_J=0.
\end{equation}

Indeed, condition (\ref{eq:rankcondk}) means that passing from $A_{-J}$ to $A$ by adjoining the $k$ columns of $A_J$ increases the rank by $k$. Therefore the images of these $k$ columns in the quotient space
\begin{equation}\label{eq:att1l}
\operatorname{span}(A)/\operatorname{span}(A_{-J})
\end{equation}
are linearly independent. Equivalently,
\begin{equation}\label{eq:directsum}
\operatorname{span}(A_J)\cap \operatorname{span}(A_{-J})={0}.
\end{equation}

Now (\ref{eq:att1k}) can be written as
\begin{equation}\label{eq:blockeq}
A_{-J}x_{-J}+A_Jx_J=0.
\end{equation}
Hence
\begin{equation}\label{eq:att1m}
A_J x_J=-A_{-J}x_{-J}\in\operatorname{span}(A_{-J}).
\end{equation}
But also
\begin{equation}\label{eq:att1n}
A_J x_J\in\operatorname{span}(A_J).
\end{equation}
By (\ref{eq:directsum}), this implies
\begin{equation}\label{eq:att1o}
A_J x_J=0.
\end{equation}
Since (\ref{eq:rankcondk}) implies in particular that the $k$ columns of $A_J$ are linearly independent modulo $\operatorname{span}(A_{-J})$, they are linearly independent among themselves, so
\begin{equation}\label{eq:att1p}
\operatorname{rank}(A_J)=k.
\end{equation}
Therefore $\operatorname{ker}(A_J)=\{0\}$, and thus $x_J$=0.

So every solution of (\ref{eq:att1k}) satisfies $x_J =0$, i.e., equation (\ref{eq:att1k}) uniquely determines the $k$ unknowns $x_j$.
\acknowledgments

The author is grateful to A. Yu. Kamenshchik, A. D. Shatkus, Tuck Choy, and M. Davidson for their interest in this work and valuable comments.

The author used ChatGPT for assistance with suggestions on mathematical methods and for improving the clarity and presentation of the manuscript. The author bears full responsibility for the content of this work.




\end{document}